\documentclass[sigconf]{acmart}

\usepackage{supertabular}

\AtBeginDocument{%
	}

\copyrightyear{2023}
\acmYear{2023}
\setcopyright{acmlicensed}\acmConference[FAccT '23]{2023 ACM Conference on Fairness, Accountability, and Transparency}{June 12--15, 2023}{Chicago, IL, USA}
\acmBooktitle{2023 ACM Conference on Fairness, Accountability, and Transparency (FAccT '23), June 12--15, 2023, Chicago, IL, USA}
\acmPrice{15.00}
\acmDOI{10.1145/3593013.3593999}
\acmISBN{979-8-4007-0192-4/23/06}

\begin{document}
	
	\title{Ghosting the Machine: Judicial Resistance to a Recidivism Risk Assessment Instrument}
	
	\author{Dasha Pruss}
	\affiliation{%
		\institution{University of Pittsburgh}
		\streetaddress{4200 Fifth Avenue}
		\city{Pittsburgh}
		\country{USA}
		\postcode{15260}
	}
\email{dasha.pruss@pitt.edu}

	\begin{abstract}
		Recidivism risk assessment instruments are presented as an `evidence-based' strategy for criminal justice reform -- a way of increasing consistency in sentencing, replacing cash bail, and reducing mass incarceration. In practice, however, AI-centric reforms can simply add another layer to the sluggish, labyrinthine machinery of bureaucratic systems and are met with internal resistance. Through a community-informed interview-based study of 23 criminal judges and other criminal legal bureaucrats in Pennsylvania, I find that judges overwhelmingly ignore a recently-implemented sentence risk assessment instrument, which they disparage as ``useless,'' ``worthless,'' ``boring,'' ``a waste of time,'' ``a non-thing,'' and simply ``not helpful.'' I argue that this algorithm aversion cannot be accounted for by individuals' distrust of the tools or automation anxieties, per the explanations given by existing scholarship. Rather, the instrument's non-use is the result of an interplay between three organizational factors: county-level norms about pre-sentence investigation reports; alterations made to the instrument by the Pennsylvania Sentencing Commission in response to years of public and internal resistance; and problems with how information is disseminated to judges. These findings shed new light on the important role of organizational influences on professional resistance to algorithms, which helps explain why algorithm-centric reforms can fail to have their desired effect. This study also contributes to an empirically-informed argument against the use of risk assessment instruments: they are resource-intensive and have not demonstrated positive on-the-ground impacts. 
	\end{abstract}

	
	
	\maketitle
	
	\section{Introduction}
	
	
	%
	Algorithmic decision-making in the public sector -- criminal law, policing, education, and public benefits -- is often introduced as a reform measure intended to address institutional inefficiency and problems of legitimacy \cite{porter_trust_1995}. 
	Recidivism risk assessment instruments, which infer a defendant's likelihood of rearrest or reconviction from past data, are presented as an `evidence-based' strategy for criminal justice reform -- a way of increasing consistency in sentencing, replacing cash bail, and reducing mass incarceration. In practice, however, AI-centric reforms can simply add another layer to the sluggish, labyrinthine machinery of bureaucratic systems and are met with internal resistance. 
	
	Consider the Sentence Risk Assessment Instrument, a recidivism risk assessment instrument implemented in Pennsylvania in 2020. The actuarial tool uses demographic factors such as age and number of prior convictions to estimate the risk that an individual will ``reoffend and be a threat to society'' -- that is, be reconvicted within 3 years of release from prison \cite{pennsylvania_commission_on_sentencing_adopted_2019}. It was adopted on the premise that it would help judges identify candidates for alternative sentences, despite public criticism that the tool would exacerbate racial biases in sentencing \cite{aclu_of_pennsylvania_testimony_2019, coalition_to_abolish_death_by_incarceration_proposed_2019, sassaman_pennsylvanias_2019}. Through a community-informed interview-based study of 23 criminal judges and other criminal legal bureaucrats in Pennsylvania, however, I find that judges overwhelmingly ignore the Sentence Risk Assessment Instrument, which they disparage as ``useless,'' ``worthless,'' ``boring,'' ``a waste of time,'' ``a non-thing,'' and simply ``not helpful.''
	
	Proponents and critics of risk assessment instruments alike tend to focus on the algorithms' technical aspects, such as their ability (or inability) to meet benchmarks of accuracy and algorithmic fairness, their proprietary nature, their predictive features, and their opacity. Many studies also assume, with no empirical basis, that bureaucrats such as judges, police officers, and government workers are prone to relying uncritically on predictive instruments -- which are often advisory. Finally, studies and audits of risk assessment instruments are frequently conducted without the input or expertise of the communities most affected by, and most experientially knowledgeable about, the ongoing effects of their implementation -- in the present context, communities impacted by incarceration. 
	
	This study takes a different approach to all three of these issues. It builds on the insights of previous empirical studies on the impacts of predictive technologies in the criminal legal system \cite{stevenson_assessing_2018, albright_if_2019, stevenson_algorithmic_2021, sloan_effect_2018, garrett_judging_2020}, ethnographic work on professional resistance in sociotechnical systems \cite{christin_algorithms_2017, brayne_predict_2020}, and input from community members to examine the impacts the Sentence Risk Assessment Instrument has had on judicial practice in Pennsylvania since its implementation in 2020.
	
	My study has several key findings. I show that criminal court judges in Pennsylvania overwhelmingly ignore the recommendations of the Sentence Risk Assessment Instrument, a form of professional resistance to algorithmic systems. I argue, however, that this algorithm aversion cannot be accounted for by individuals' distrust of the tools or automation anxieties, per the explanations given by existing scholarship \cite{dietvorst_algorithm_2015, brayne_technologies_2020}. Indeed, I find that even staunch supporters of risk assessment reform measures are critical of this particular tool. Instead, 
	I identify three organizational factors that jointly explain the instrument's non-use: disparate county-level norms about pre-sentence investigation reports; alterations made to the instrument by the Pennsylvania Sentencing Commission in response to years of public and internal resistance; and problems with how information is disseminated to judges. My qualitative analysis thus provides an explanation of the Pennsylvania Sentencing Commission's own initial data analysis that the tool has had no impact on sentencing \cite{pennsylvania_commission_on_sentencing_commission_2021}, the inconsequential outcome of a decade-long process to satisfy a 2010 state legislative mandate for a sentencing risk assessment instrument. I also note two potential unexpected consequences of the tool's adoption: additional hidden labor for the probation department and longer pre-trial detention times for defendants. 
	
	These findings shed new light on the important role of organizational influences on professional resistance to technology, which helps clarify one reason that algorithm-centric reforms can fail to have their desired effect. This study thus lends empirical support to a practical argument against the use of risk assessment instruments: they are resource-intensive and have not demonstrated positive on-the-ground impacts.

	\section{Background and Related Work}
	
	\subsection{Risk Assessment Instruments and Human Discretion}
	
	Scholarship on predictive technologies in the public sector has exploded in recent years \cite{chouldechova_fair_2017, brown_toward_2019, fogliato_impact_2021, levy_algorithms_2021, akpinar_effect_2021,  meyer_flipping_2022}. The use of algorithmic decision-making in the criminal legal system has been particularly controversial, with reason. 
	The claim that risk assessment instruments promote progressive criminal justice goals in practice is largely speculative -- the few existing empirical studies suggest that risk assessment tools have had little to no impact \cite{stevenson_assessing_2018, sloan_effect_2018, garrett_judging_2020, stevenson_algorithmic_2021} -- and 
	a vocal chorus of critics has stressed that such instruments could exacerbate racial disparities in pretrial, sentencing, and parole decisions because they base predictions on (and reproduce) structurally racist patterns in the US criminal legal system \cite{harcourt_against_2008, hannah-moffat_actuarial_2013, angwin_machine_2016}.
	
	To be sure, algorithmic bias is worth addressing seriously and can be reason alone to condemn the use of a particular instrument. But a key detail often neglected in discourse about risk assessment instruments and other public sector algorithmic systems is that their recommendations are advisory. 
	%
	Algorithmic systems 
	are socially situated, interacting and entangling by necessity with people, institutional practices, and societal norms \cite{alkhatib_street-level_2019, mittelstadt_ethics_2016, pruss_mechanical_2021-1, glaser_biography_2021}.  Individuals like judges and police officers make on-the-ground discretionary decisions -- what Michael Lipsky refers to as `street-level bureaucracy' \cite{lipsky_street-level_1980} -- that ultimately impact the lives of individual people, not the technical details of the algorithmic instruments on their own, and human judgment can interact with algorithmic decision-making systems in unexpected ways. The few studies of how risk assessment instruments are actually used have shown that judges differ widely in their adherence to recommendations and follow them inconsistently for different types of defendants \cite{stevenson_assessing_2018, garrett_judging_2020, stevenson_algorithmic_2021}.
	
	For example, human decision-makers can selectively follow algorithmic recommendations to the detriment of individuals already likely to be targets of discrimination. In Kentucky, a pretrial risk assessment tool -- intended as a bail reform measure -- increased racial disparities in pretrial releases and ultimately did not increase the number of releases overall because judges ignored leniency recommendations for Black defendants more often than for similar white defendants \cite{albright_if_2019}. Likewise, judges using a risk assessment instrument in Virginia sentenced Black defendants more harshly than others with the same risk score \cite{stevenson_algorithmic_2021}.
	
	In other contexts, human discretion can correct for algorithmic bias. In Pennsylvania, a recent study about racial bias in an algorithm that screens for child neglect showed that call screeners minimized the algorithm's disparity in screen-in rate between Black and white children by ``making holistic risk assessments and adjusting for the algorithm's limitations'' \cite{cheng_how_2022} (see also \cite{de-arteaga_case_2020}).  Virginia's risk assessment instrument would have led to an increase in sentence length for young people had judges adhered to it; however, because judges systematically deviated from recommendations, some of the instrument's potential harms (and benefits) were minimized \cite{stevenson_algorithmic_2021}.
	
	Of course, another way that human discretion can interact with algorithms is not to interact with them. Algorithm aversion -- the reluctance to follow algorithmic recommendations -- is thought to arise from lack of confidence in algorithmic systems \cite{dietvorst_algorithm_2015}; however, experimental research on algorithm aversion has focused on individual and algorithm factors, neglecting the role of social context and organizational factors \cite{mahmud_what_2022}. Sociological work shows that resistance to algorithms happens in contexts where individuals feel that their agency or power is being threatened by a new technology, as illustrated by Sarah Brayne in her ethnography of LAPD officers using PredPol, as well as by Ang\`{e}le Christin in her ethnography of prosecutors and judges using a pretrial risk assessment instrument \cite{brayne_predict_2020, christin_algorithms_2017}. Police officers and legal professionals alike felt threatened by how these new technologies could be used to surveil their performance and limit the role of their discretion, resulting in professional resistance to algorithmic systems in the form of adversarial data obfuscation -- the process of manipulating a system's data to make it useless -- and foot-dragging. 
	
	These dynamics can also intersect. In Virginia, judges had highly divergent attitudes toward (and literacy about) risk assessment and varied widely in whether and how they adhered to algorithmic recommendations \cite{garrett_judging_2020}. Understanding how  these possible forms of human-algorithm interaction apply in a given case thus requires not only empirical research in a context of application but also attention to the social and organizational factors at play.
	
		\begin{figure*}[h!]
		
		\includegraphics[width=\textwidth]{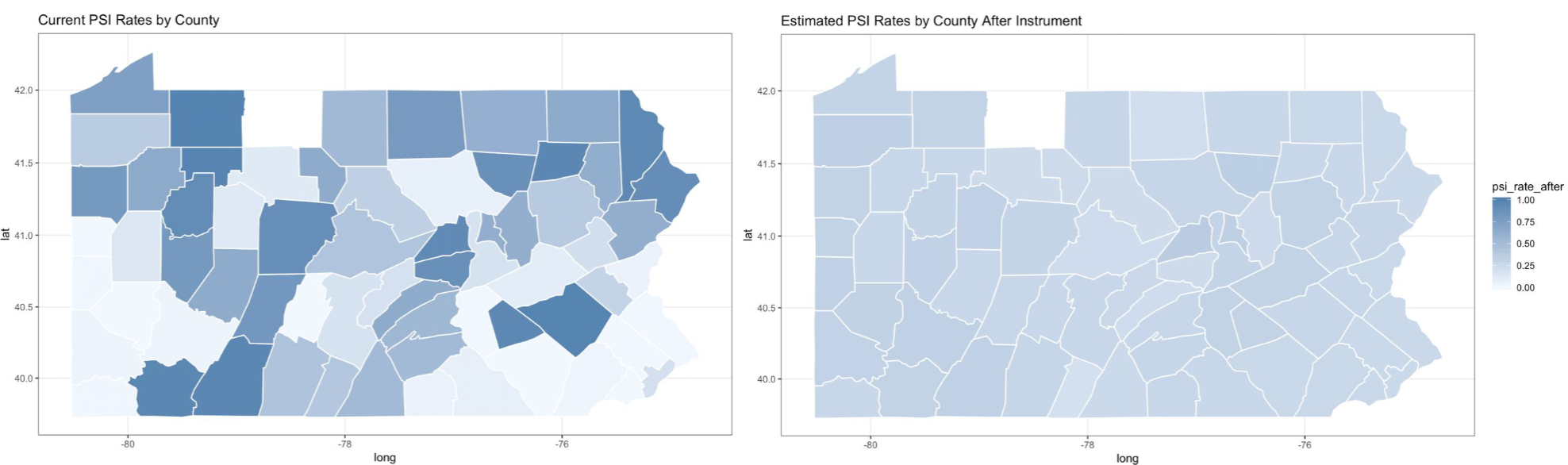}
		\caption[``Comparison of PSI Rates Before and After the Instrument'']{\small``Comparison of PSI Rates Before and After the Instrument,'' a figure from a third-party audit of the tool, which states that ``if PSIs were to completed [sic] following the rate at which the instrument identifies high- or low-risk offenders, the PSI rates across counties will be more consistent'' \cite{becerril_validation_2019}. 
		}
		\label{fig:psi_rates_2}
	\end{figure*}
	
	\subsection{The Sentence Risk Assessment Instrument}
	
	Amid the chaos of the early months of the pandemic, criminal courts throughout Pennsylvania were instructed to begin consulting the Sentence Risk Assessment Instrument when sentencing crimes, with the aim of helping judges identify candidates for alternative sentences. The instrument applies to non-DUI defendants being sentenced following an open plea or trial, July 2020 onward. It generates a risk score of an individual's risk of recidivism based on demographic factors including age, gender, number of prior convictions, current conviction offense type, and prior juvenile adjudication. These variables are given weights, depending on their degree of association with reconviction. Young age, for example, is associated with recidivism; individuals under the age of 21 receive 5 points and those over 49 receive 0 points. These points are summed to make the risk score, which ranges from 0 to 18 and is binned into low-, typical-, and high-risk categories.

	The tool recommends seeking `Additional Information', typically a pre-sentence investigation report (PSI), for individuals with a low or high risk of recidivism ``for whom additional information may assist the court in determining candidates for alternative sentencing'' \cite{pennsylvania_commission_on_sentencing_adopted_2019}. The instrument is thus intended to influence a judge's decision to order a PSI for a given criminal defendant, with the presumption that information contained within PSIs will in turn influence a judge's decision to assign an alternate sentence. Currently, PSI-ordering rates in Pennsylvania vary substantially county-to-county, as do the contents of the reports; one of the expected outcomes of the tool's adoption was thus to minimize county-level disparities in how often, and for which kinds of defendants, judges choose to order a PSI (\autoref{fig:psi_rates_2}).

	The Pennsylvania Sentencing Commission -- a legislative agency that advances ``fairer and more uniform decisions at sentencing, resentencing, and parole'' -- was tasked with fulfilling a 2010 state legislative mandate to develop the instrument \cite{pennsylvania_commission_on_sentencing_pennsylvania_2022}. However, the Commission's members soon found themselves embroiled in controversy. From 2017--2019, the Commission received over 100 overwhelmingly negative public testimonies about the tool from sources including AI Now, the ACLU, high-profile academics, and local community organizations. Critics argued that the ``racist tool'' \cite{aclu_of_pennsylvania_testimony_2019} could ``perpetuate the racial biases and stigmas inherent in our criminal legal system'' \cite{coalition_to_abolish_death_by_incarceration_proposed_2019}. The instrument also met intense criticism from within the criminal legal system, particularly from probation officers, who argued that the Sentence Risk Assessment Instrument was ``an unnecessary burden in time, effort, and resources'' and would increase the workload of ``already overwhelmed'' county probation departments \cite{county_chief_adult_probation_and_parole_officers_association_of_pennsylvania_re_2019}. 
	
	In informal interviews, Commission staff explained that they were legally required to implement the legislative mandate despite these criticisms, lamenting that ``from the start ... there has been no public support for the development and use of risk assessment at sentencing'' \cite{pennsylvania_commission_on_sentencing_adopted_2019}. 
	Commission staff, to their credit, engaged with the public through a transparent and iterative process of development, removing piece by piece the most controversial parts of the instrument and seeking further public comment each time. For instance, an earlier version of the instrument showed judges not only an individual's risk score but also a detailed risk distribution, indicating to the judge exactly where the defendant's numerical score falls relative to other individuals. The final version of the tool only shows judges a small text box with the words ``Additional Information'' (if the defendant is low- or high-risk), or ``NA'' (if the defendant is moderate-risk). The Commission also changed its outcome variable from rearrest to reconviction in response to public testimonies, which argued that arrest is not only a poor predictor of actual crime but also racially correlated due to racial profiling by police \citep{sassaman_testimony_2018}. The core concerns of the public and probation officers, however, went unaddressed -- the tool was still implemented, and no additional resources were allocated to assist probation departments with the anticipated increase in ordered reports.
	
	As part of the state's Evidence-Based Practices Strategic Plan, the Commission solicited an external review of the tool by Carnegie Mellon University researchers in 2019. This audit focused on technical benchmarks of validity, accuracy, and fairness, and made several recommendations, including discarding the high risk category due to low accuracy; removing gender as a predictive factor; raising the high-risk category cutoff to increase its accuracy; and not deploying the violent crime risk scale component of the instrument due to an unacceptable level of false positives \cite{becerril_validation_2019}. The Commission voted to follow the latter two recommendations. Notably, the audit does not consult relevant stakeholders or mention the tool's interactional effects with judicial discretion or other social factors, instead including projections (e.g., \autoref{fig:psi_rates_2}) that assume complete uptake of the tool. Later in 2019, the Commission voted to adopt the tool, and the Sentence Risk Assessment Instrument was formally rolled out in July 2020.
	
	\section{Methods}
	
	
	To understand how judges use and interpret the recommendations of the Sentence Risk Assessment Instrument, I conducted semi-structured interviews with 15 criminal court judges \cite{merriam_qualitative_2015}, as well as unstructured interviews with three probation officers and four current and former Pennsylvania Sentencing Commission staff. \vspace{.5em}

	\noindent \textbf{Community Recommendations.} Drawing on standpoint theory \cite{harding_rethinking_1992}, I hired two justice-impacted individuals from the community organization Coalition to Abolish Death by Incarceration (CADBI) as consultants on the project in an effort to prioritize the affected community's interests and knowledge in developing my interview questions. One of the consultants was formerly incarcerated and the other works supporting incarcerated people and their families. Prior to conducting interviews, I met with both consultants to determine the scope of the project's research questions and later solicited their written and verbal feedback on a draft of an interview guide I produced based on this initial meeting; I compensated consultants for their time at a rate of \$40/hour. One individual expressed concern that the new risk assessment tool would make judges more likely to ignore the humanity and personal circumstances of the people they sentenced and suggested gauging judges' awareness of this issue. Consultants also wanted to include interview questions about the personal nature and impacts of their sentencing decisions. Based on this feedback, I added questions to the interview guide to probe judges' concerns about the instrument and which personal factors judges consider in their sentencing decisions. \vspace{.5em}

	\noindent \textbf{Recruitment and Demographics.} I conducted interviews with judges from Allegheny, Philadelphia, Delaware, Dauphin, and York Counties. In each county, I initially recruited judges through emailed and physically mailed study invitations and follow-up phone calls regarding these invitations until I received a response or the timeframe for my data collection passed. Other judges were recruited through snowball sampling from initial responders. 
	I made an effort to select a sample of judges with variation \cite{weiss_learning_1995} across county, political orientation, favorability to risk assessment instruments, age, gender, race, and time served as a judge (see \autoref{table:demographics} for the results of a demographic survey given to interviewed judges; see \autoref{sec:survey} for the survey). Nevertheless, it is likely that the sample over-represents judges with higher-than-average familiarity with risk assessment instruments, since these individuals are more likely to agree to an interview about such instruments and in turn likely to refer study participants similar to themselves \cite{parker_snowball_2019}. I continued recruiting and interviewing judges until I achieved saturation, that is, I no longer heard new information in my interviews \cite{small_how_2009}. In total, I attempted to recruit 86 judges, resulting in a response rate of 17\%. 	\vspace{.5em}
	
	\begin{table}[t!]
		\small
		\centering
		
		\begin{tabular}{  l c r  } 
			
			\textbf{Sex} & \textbf{Frequency} & \textbf{\%} \\ 
			
			\hline

			Male & 8 & 53.3\%\\
			Female& 7 & 46.6\%\\
			\textbf{Age}\\ 
			40--49 years & 1 & 6.7\% \\
			50--59 years & 6 & 40\%\\
			60--69 years & 5 & 33.3\%\\
			70--79 years & 2 & 13.3\%\\
			No response & 1 & 6.7\% \\ 
			
			\textbf{Years as a judge}\\
			0--2 years & 2 & 13.3\%\\
			2--5 years & 2 & 13.3\%\\
			5--10 years & 5 & 33.3\%\\
			10--20 years & 5 & 33.3\%\\
			20+ years & 1 & 6.7\%\\
			No response & 1 & 6.7\% \\

			\textbf{Race/Ethnicity} \\
			White or Caucasian & 11 & 73.3\%\\
			Black or African American & 3 & 20\%\\
			No response & 1 & 6.7\%\\
			
			\textbf{County} \\
			Allegheny & 4 & 26.7\%\\ 
			Philadelphia & 5& 33.3\%\\
			Dauphin & 2& 13.3\%\\
			Delaware & 2 &13.3\% \\
			York & 2& 13.3\%\\
			
			\textbf{Political Orientation} \\
			Democrat & 7 & 46.7\%\\
			Republican & 4 &26.7\%\\
			Non-Partisan/Independent & 2 & 13.3\%\\
			No response & 1 & 6.7\%\\

			\hline
		\end{tabular}
		\vspace{.8em}
		
		\caption[Features of interviewed judge population based on demographic survey]{Features of interviewed judge population based on demographic survey (15 judges).}
		
		
		\label{table:demographics}
	\end{table}

	\noindent \textbf{Interview Process.} Interviews with judges ranged from 30 minutes to 2 hours, with a median length of 50 minutes, and were conducted over video call, by phone, and in person; follow-up questions were answered over email and follow-up interviews were conducted with three judges. Interview topics included the career trajectories of judges; sentencing practices; training, impressions, and use of the risk assessment tool; and attitudes about risk assessment instruments more broadly (see \autoref{sec:interview_guide} for interview questions). Interviews with probation officers and Sentencing Commission staff were unstructured and helped triangulate interview data from judges and inform the research project more broadly. This study received an IRB exemption and I made sure not to include any information from interviews that might contain identifying information in order to keep the identities of study participants anonymous. \vspace{.5em}
	

	\noindent \textbf{Qualitative Analysis.} I produced an analytic memo for each interview \cite{miles_qualitative_2014}, reviewed interview transcriptions generated by OpenAI's Whisper 2-3 times, and relistened to audio recordings twice. I coded interviews iteratively to identify and label repeating ideas in the interviews, moving between inductive coding and data collection to refine themes and look for disconfirming evidence as further interviews were conducted \cite{miles_qualitative_2014}. I converged on seven high-level themes, each with 4-10 sub-themes:  
	sentencing practice; PSI ordering behavior; information and training about the tool; familiarity with and misconceptions about the tool; use of the tool; desires and concerns about the tool; and attitudes about risk assessment instruments more broadly (see \autoref{sec:code_table} for a code table). In order to ensure internal validity, I used member checks and triangulated data from from multiple sources \cite{merriam_qualitative_2015}, including participant observations with chambers and courthouse staff during two in-person site visits to the Allegheny and Philadelphia county courthouses, public testimony documents, instrument development documentation, and recorded meetings of the Pennsylvania Sentencing Commission, all of which are publicly available on the Commission's website. \vspace{.5em}
	
	\noindent \textbf{Positionality.} As a white woman, an academic researcher, and a regular contributor to activist initiatives opposing the use of carceral technology in my local community, I acknowledge that my positionality shaped the research questions I was interested in pursuing as well as my interactions with interviewees. I have participated in rallies and other events organized by the community organization I collaborated with, which helped me build rapport with my community consultants despite my privileged academic position, race, and lack of personal contact with the criminal legal system; nevertheless, these differences likely shaped the feedback my consultants were comfortable giving me. On the other hand, my privileged position as a white researcher from a respected local university helped me access and build rapport with judges, many of whom were also white and received their legal training at elite academic institutions. 
	
	\section{Results}
	
	With respect to tool uptake, I rapidly achieved saturation in my findings: judges were not interested in, and did not consult, the Sentence Risk Assessment Instrument. Only two of the judges I spoke with reported regularly consulting the instrument, and even these individuals could not recall a single instance in which it had affected their decisions to order a PSI. In more populated counties (Allegheny and Philadelphia), I noticed repeating data by my third interview; I continued getting the same result from judges in smaller counties (Dauphin, Delaware, York), where political orientation and PSI-ordering behavior differed from the larger counties, which I expected to correlate with tool use. However, regardless of county size, judges almost unanimously did not use the instrument. This finding is further supported by the Pennsylvania Sentencing Commission's own quantitative analysis of the tool, which shows that there was no requisite change in PSI-ordering rates after the implementation of the tool. Moreover, my study sample likely over-represents individuals with atypically high interest in and knowledge about the tool; the fact that even these judges ignore the tool supports the generalizability of my finding.
	
	Although I achieved saturation with respect to lack of tool uptake, I saw a wide range of responses for my other interview themes, especially PSI-ordering behavior, familiarity with and misconceptions about the tool, desires and recommendations about the tool, concerns about the tool, and attitudes about risk assessment instruments more broadly. That is, I saw variety in the reasons \textit{why} judges ignored the tool. Here I present the main ones. 
	
	\subsection{``I find it to not be particularly, um... helpful.''}
	\label{sec:helpful}
	
	The most common reason that judges did not use the tool was that they simply did not find it useful. This is due in part to the work of activists, lawyers, and academics who, over years of public testimony hearings, successfully pressured the Pennsylvania Sentencing Commission to remove the most controversial parts of the instrument, including directly showing judges risk scores and detailed risk distributions. The implemented version of the tool recommends ordering additional information about low- and high-risk defendants, in keeping with the original goal of helping judges identify candidates for alternative sentences. However, none of the judges I spoke with were looking to change their PSI-ordering behavior. Judges reported either ordering PSIs for all trial cases, ordering PSIs for more serious trial cases, or almost never ordering PSIs; this behavior reflected how useful judges found the PSIs themselves, whose contents vary by county. Nearly half of the judges I talked to also did not find the contents of PSIs helpful because in many counties, including the state's most populous Philadelphia and Allegheny counties, the reports contain information judges can get simply by talking to the defendant. In other words, the tool intervenes on a factor -- PSI-ordering behavior -- that judges are uninterested in changing, and falsely assumes that successfully influencing PSI-ordering behavior will in turn influence sentencing decisions. 
	
	Five judges explicitly used the words ``useless'' or ``worthless'' (sometimes with an expletive) to describe the Sentence Risk Assessment Tool. Over half of the judges also stated that they would have preferred to see different information presented to them at sentencing time, including the causal impacts of different sentencing practices on recidivism, a risk and needs responsivity risk assessment, information about how the risk assessment was derived (``Show me the math''), and information about risk categories (``It would be better if they said high, moderate, or low, to be honest''; one judge said they would only want to see information about low-risk defendants, while another said they would only want to see information about high-risk defendants). 
	
	\subsection{``I have no idea where it is on the form; I don't recall looking at it at any point.''}
	
	Another common reason that judges ignored the tool, which often overlapped with judges' perceptions of the tool's uselessness, was a lack of information about what the tool did or what form its recommendation appeared on. As one judge put it, ``I never knew where that information was going to be provided for me. Was it going to come in an email? A news blog? A winter weather alert? I had no idea.''  Several judges explicitly asked me to show them where on the sentence guideline form that judges routinely receive at sentencing time -- ``the world's least user-friendly form'' -- the recommendation appears. Another judge called their supervising judge during my interview because they did not believe me that a sentencing risk assessment instrument was in use in their county. With two exceptions, every judge I spoke with revealed some degree of misconception about the tool during the course of my interview, such as the claim that the tool shows judges risk scores (it does not), that the tool applies to DUI cases (it does not), and that the judge has to do something in order to generate the risk assessment (they do not; it is automatically generated and appears on the sentence guideline form). Several interviewed judges were ashamed about being on the record about their lack of awareness of the tool, while others used their lack of knowledge about the tool as a reason to decline participation in my study. 
	
	
	Nearly all judges had low literacy of the tool, despite the Commission's claim that, effective January 1, 2020, it would ``conduct a six-month training and orientation for judges and practitioners related to the use of the Sentence Risk Assessment Instrument, the purpose of the recommendation, and the type of information recommended'' \cite{pennsylvania_commission_on_sentencing_adopted_2019}. Many judges and probation officers remarked that the tool -- and how to use it -- had been poorly publicized. In personal conversations, Commission staff explained that their information campaign had been derailed by the start of the pandemic coinciding with the roll-out of the tool.  
	
	More broadly, however, my findings indicate systemic problems with how information is disseminated to judges in Pennsylvania. In one particularly revealing moment, a judge told me that they were attending a virtual Continuing Judicial Education session over video call in the background of their computer -- during our interview. The problem of judicial education was echoed to me by a chief probation officer, who lamented that even with respect to the risk assessment already included in PSIs in their county, the probation department had not done much in the way of educating judges about how to interpret risk assessment information, adding that many judges ``didn't really understand how it applies to the work that they do'' and that this was likely the case statewide.\footnote{This issue extends beyond Pennsylvania; low literacy about risk assessment among judges has also been documented in Virginia \cite{garrett_judging_2020}.} 
	
	\subsection{``It's unworkable. I don't know how you're building that into numbers.''}
	\label{sec:unworkable}
	
	In addition to misinformation and perceptions of uselessness, skepticism or concern about risk assessment instruments more broadly was often a complementary reason that judges cited for ignoring the Sentence Risk Assessment Instrument, though it was typically a secondary issue. These concerns fell roughly into three categories. 
	
	The most common concern, which roughly half of judges expressed, was that the tool ignored a defendant's humanity. Notably, this was a central issue raised by CADBI members in their feedback on my study design; one formerly incarcerated individual worried that the new risk assessment tool would make judges more likely to ignore the humanity and personal circumstances of the people they sentenced. ``Each individual has a history that brought them to this space,'' this consultant told me. ``There must be individualization.'' Judges echoed this point, raising concern about ``having a formula that takes away my ability to see the humanity of the people in front of me''; another judge argued that ``cookie cutter justice doesn't work'' and that risk assessment was ``merely labeling and boxing''; a third said, ``I don't know how you can reduce all of the human factors that go into, you know, sentencing or making a bond decision and, and put it into a number, you know, I just, I just think that there are a world of human factors that need to be considered.'' These judges emphasized the crucial role that individual narratives and personal context played in their sentencing decisions.  Most judges also indicated that they did not assign central importance to aggregated recidivism risk in their sentencing decisions (with the exception of recidivism risk for sex crimes). Rather, they were interested in the personal trajectories of criminal defendants, particularly escalation toward violent behavior; whether a defendant was employed; and drug use.
	
	Another common concern judges raised was about the tool's bias, especially racial bias. One judge, who identified as Black, was critical of the discriminatory potential of the tool: ``Who's making the determinations? Who's interpreting the statistics? You can say anything with statistics.'' Another judge noted the third-party audit's finding that the tool's high-risk category was less accurate than the low-risk category, commenting that this could be ``prejudicial to certain minority groups because there was an historically higher arrest rate, possibly related to things like race rather than actual criminal activity.'' Judges were concerned about other biases as well -- a judge who was otherwise an advocate of risk assessment tools claimed that the tool was biased in favor of sex offenders (a claim that is not factually accurate), while two others commented that age was an unfair indicator of recidivism because minorities are statistically more likely to be stopped by police at a younger age. This concern about bias was not unanimously shared, however; other judges acknowledged that the tool had biases but maintained that these were still better than human biases: ``You can never take all biases out. You can never take out -- there's biases, people get arrested -- what's in it, but you can continue to work on the tools to try to make them as fair as possible. But it's better than individuals.'' One judge even claimed that ``[risk assessment tools] have been deliberately distorted as being racist, as being not accurate, as being using wrong statistics and things like that.''
	
	The third most common concern was that the tool was worse than the discretion of experienced judges.  
	A common refrain from judges was that younger, less experienced judges might get more benefit from the risk assessment tool, but that for more experienced judges, such an instrument was unnecessary. There was also a general sentiment from judges that personal discretion was a centrally-defining feature of what it means to be a judge; one judge with over a decade of experience firmly announced in the first 10 seconds of our conversation that they were ``elected to be a judge, not a robot.'' Nine judges independently brought up that judges ``don't want to be told by anybody what to do;'' however, those same judges did not view themselves as being in this category. Seven judges said their own sentencing practice was better than other judges, describing their sentencing using adjectives like ``different,'' ``atypical,'' or (pleasantly) ``shocking'' to defendants. Several judges were critical of any efforts to limit their discretion, including sentencing guidelines, which are supposed to standardize sentence lengths based on an individual's prior record score and the gravity of their current offense. One judge aptly summarized this particular concern: 
	%
	``[The legislature] is trying to give us more narrow options on what we can do. And, and I don't like that, because I think that there's a reason that we're up there -- we're up there because supposedly we've demonstrated some ability to think more broadly about the whole system and to make a better decision than just something that's electronically generated. You know if you're going to do it all based on a computer program, then you don't need me out there.''

	Importantly, however, judges' skepticism about risk assessment instruments should not be conflated with skepticism toward data-driven strategies in criminal justice more broadly. As already mentioned, many judges reported wanting access to more data at sentencing time -- just not the kind of information provided by this risk tool. Moreover, most judges did in fact acknowledge the importance of consistency in sentencing and, with few exceptions, reported complying with sentencing guidelines. With the exception of two judges, the skeptical claims above were regularly expressed alongside pro-data and pro-science stances at other points within the same interview. One judge, who had expressed concerns about the tool's racial bias earlier in our interview, maintained that ``I'm a believer in science. This [risk assessment] is science, so we need to use it.''  
	
	\subsection{``Anything that slows down processing will be met with resistance.''}
	\label{sec:resistance}
	
	Several judges worried that the Sentence Risk Assessment Instrument could have unexpected downstream consequences, were it to be used. The Commission ``expressly disavows the use of the sentence risk assessment instrument to increase punishment'' \cite{pennsylvania_commission_on_sentencing_adopted_2019}. However, as one judge and public testimonies pointed out, judges can still infer risk levels from the `Additional Information' label, and empirical evidence from other states suggests that judges are more likely to use risk information to detain individuals longer \cite{human_rights_watch_not_2017}. ``People know it's called the risk tool. If it's `Additional Information', there may be some concern about how dangerous the defendant is,'' a judge noted. Moreover, if judges followed the tool's recommendation to order PSIs for low-risk defendants -- who often have minor sentences -- then the tool could have the unintended effect of detaining these defendants longer pre-trial, since ordering a PSI can take 60 days or longer, depending on the county. Another judge remarked, ``I'm not letting them [the defendant] sit 8 more weeks in jail because some computer program said so.'' 
	
	Speaking about unintended impacts in other parts of the criminal legal system, two probation officers also shared worries about the tool creating unnecessary -- and invisible -- labor for their departments, which are tasked with generating the risk and needs responsivity assessments that go into the PSIs in some counties. One of these officers said they feared they were going to get ``a flood of cases'' where judges were ordering PSIs, but that ``thankfully that has not happened'' because they did not have the resources to handle such a surge. They said they would like to see the system someday permit having such an assessment done for every defendant, but that this would ``require a lot of resources, a lot of resources.'' The second officer raised the concern that the tool, if widely used, would ``significantly slow down'' the already-backlogged sentencing process, which they said could cause individuals to spend even more time awaiting trial in jail. To this probation officer, the risk assessment instrument was just ``another unfunded mandate, the burden of which was going to fall on county probation.''

	\subsection{``We're past that train stop and a little bit further down the tracks.''}
	
	A minority of judges I spoke with were knowledgeable, vocal advocates of other risk assessment instruments and the Pennsylvania Sentencing Commission's other projects. Even among these four judges, however, only one claimed to be regularly consulting the Commission's tool, with the caveat that it had never changed their PSI-ordering behavior. Judges in this group were either advocates of using risk assessment at other stages of the criminal legal pipeline, such as at preliminary arraignment, or were serving in counties where the Ohio Risk Assessment Instrument (ORAS), a significantly more detailed risk and needs responsivity risk assessment, is conducted by the probation department and is already a routine part of the PSIs that judges receive. One self-described ``cheerleader'' for risk assessment instruments explained: ``I like the [Commission's] tool, I just like our tool [the ORAS] better -- it's shinier and faster.'' This was the position of two of the probation officers I spoke with as well.
	
	In sum, although nearly all of the judges I interviewed reported ignoring the Sentence Risk Assessment Instrument, their reasons for this varied. This suggests a nuanced explanation for aversion to algorithmic systems in the criminal legal system that is neglected in existing discussions that are centered largely around lack of confidence in technology and fears of deskilling and surveillance. The rest of this paper discusses the implications of this finding for scholarship on algorithmic resistance and risk assessment instruments.
	
	\section{Discussion}

	`Evidence-based' sentencing strategically positions the objectivity and accuracy associated with algorithmic decision-making systems as a solution to institutional crises of mass incarceration and inefficiency.\footnote{For discussions of the relationship between quantification, objectivity, and scientific authority, see \cite{porter_trust_1995, galison_algorists_2019} and \cite{espeland_accountability_2007} for a discussion of quantification in law specifically.} But the few existing on-the-ground studies of risk assessment instruments -- this study included -- show that the tools' impacts are different than what either critics or proponents had anticipated. One reason for this is that, much like any institutional reform, the success of algorithm-centric reforms is contingent on the organizational conditions in which they are introduced. An algorithm that is intended to assist decision-makers but is developed without attention to their actual needs, or whether and how they will actually use it, is unlikely to have the anticipated effect, and whatever effect it does have will vary by individual. An algorithm that intervenes on a locus -- PSI-ordering -- that is highly variable by county and a largely settled behavioral pattern is unlikely to alter that behavior. An algorithm whose success relies on the effective dissemination of information in an institutional context in which judges can be interviewed at the same time as attending virtual training sessions is unlikely to have an effect. Crucially, none of these statements have anything to do with the algorithm's bias or accuracy, which are typically the focus of algorithmic audits and one of the main criticisms of risk assessment instruments. 
	
	The implications of this study can thus be distilled into two main points: an understanding of resistance to technology that considers organizational factors is better able to capture real-world cases of algorithm aversion; and empirical research on the inefficacy of risk assessment instruments supports an alternative argument for their abolition. I discuss these in turn. 
	
	\subsection{Algorithm Aversion from an Organizational Perspective}
	
	As one judge aptly summarized it, ``there was a lot of resistance to the tool'' -- not only from the community but also from public defenders, probation officers, criminal attorneys and, as I have shown, judges themselves. A standard algorithm aversion explanation for this could be individuals' lack of confidence in the tools \cite{dietvorst_algorithm_2015}. 
	Distrust is, no doubt, an important part of the story, particularly with respect to public resistance to the instrument. But does lack of confidence explain the resistance from judges? As I discussed in \S\ref{sec:unworkable}, some judges did cite lack of confidence in the instrument's predictions as one reason for not wanting to use them; however, this was not the primary reason but rather something that came up later in the interview once I started probing about their other concerns about the instrument. 
	Moreover, a weak majority of judges I spoke with were supportive of using risk assessment instruments in some capacity -- if not at sentencing, then at some other stage of the criminal legal pipeline -- and often reported wanting \textit{more} empirically-derived information to assist decision-making (\S\ref{sec:helpful}). In general, I observed among most judges a strong pro-data mentality. In short, lack of confidence is a simplified, algorithm-centered explanation that does not provide an adequate explanation of this real-world case -- such as why judges who are self-avowed ``cheerleaders'' of other risk assessment instruments used in their counties are still critical of the Commission's tool. 
	
	Brayne and Christin provide another, sociological explanation: judges may be engaging in behavior like foot-dragging due to fears of deskilling and managerial surveillance \cite{brayne_technologies_2020}. In their studies of predictive policing and pretrial risk assessment, Brayne and Christin found resistance to algorithms to be strongest in cases of function creep, where algorithmic tools served the added purpose of increasing managerial control and surveiling bureaucrats' productivity. While this sort of function creep is, for now, absent in the present case, I did see some evidence of automation anxieties and fears of deskilling. Some judges -- particularly in Philadelphia -- expressed antagonism toward any mandates that were intended to limit judicial discretion, including sentencing guidelines. Almost all the judges described their own discretion as a strength, not a weakness, though they were also often critical of other judges practicing the wrong kind of discretion, and reductions in judicial discretion were sometimes perceived negatively -- recall the judge who compared using risk assessment instruments to being ``a robot'' (\S\ref{sec:unworkable}). 
	Despite their concerns, however, most judges still expressed agreement with the premise of data-driven sentencing, and few opposed the use of some discretion-limiting measures, such as sentencing guidelines. This makes it unlikely that judges' resistance to the risk assessment instrument is entirely ``fueled by fears of deskilling and heightened managerial surveillance'' \cite{brayne_technologies_2020}.\footnote{Brayne and Christin also propose the thesis that predictive technologies displace discretion to less visible areas within organizations. I found some unexpected support for this thesis in counties where an additional risk and needs risk assessment instrument is included in PSIs. Judges revealed to me that probation officers have an enormous amount of discretion in how they prepare such reports; in one county, PSIs even include concrete recommendations for what an individual's sentence should be -- a determination made by the probation officer preparing the report. 
	}
	
	In this case, a more adequate explanation for why judges ignore the tool has to do with the organizational influences that led to the tool's development, policies about the contents of PSIs, and problems in how information is disseminated to judges. One probation officer described the Pennsylvania Sentencing Commission as ``trying to make certain groups happy'' -- that is, the public, the legislature, the judges, and probation officers. One of the outcomes of this negotiation process was the selection of a less-publicly-controversial locus of intervention for the tool: the decision to order a PSI. 
	
	However, judges did not report PSIs influencing their sentencing decisions except in very unusual situations, such as where a criminal record is stale.\footnote{In such cases, they may use the background information to go below sentencing guidelines.}  Typically, judges said that PSIs are not very helpful and never ``dramatically changed [their] mind'' about a sentence; this was the case even for PSIs that contained the more detailed risk assessment. This means that the final version of the tool is, at best, useless for judges; not using the tool was largely a response to this fact, complemented by widespread low literacy about the tool. At worst, judges' adherence to the tool's recommendations could have produced `ghost work' \cite{gray_ghost_2019} for probation departments and detained individuals longer pre-trial (\S\ref{sec:resistance}). But activists still see the final weakened tool as a win. As Hannah Sassaman, a Philadelphia-based community organizer, told news outlets the day after the tool's adoption, ``the tool that the Commission instituted yesterday was massively changed over the past few years from one that actively centered racist factors in guessing the future of a sentenced person, to one that will be considerably less damaging'' \cite{gross_pennsylvanias_2019}.

	\subsection{A Resource Argument Against Risk Assessment Instruments}
	
	A defender of evidence-based sentencing could make the case that, had the Sentence Risk Assessment Instrument (or the legislative mandate it was built to satisfy) been designed differently -- and had the public been more receptive to its development -- then perhaps judges would not have been so resistant to it, and perhaps more people would have gotten alternatives to prison sentences as a result. But empirical research on risk assessment instruments used in other states -- mostly for pretrial detention decisions -- suggests that the tools' impacts have been minimal, unfairly distributed, and have tended to wash out over time \cite{stevenson_assessing_2018, sloan_effect_2018, albright_if_2019, garrett_judging_2020, stevenson_algorithmic_2021}. This has been the trend even for tools with greater uptake and more significant loci of intervention than the decision to order a PSI. Empirical research also suggests that risk assessment instruments introduce an element of arbitrariness to decision-making, such as sharp differences in sentencing decisions for individuals with risk scores that fall near the low-risk category cutoff \citep{stevenson_algorithmic_2021}. As economist and legal scholar Megan Stevenson starkly puts it, ``Somehow, criminal justice risk assessment has gained the near-universal reputation of being an evidence-based practice despite the fact that there is virtually no research showing that it has been effective'' \cite{stevenson_assessing_2018}.  
	
	This research thus contributes another case study to an alternative, empirically-informed argument for abolishing recidivism risk assessment instruments: in practice, these algorithm-centric reforms have no significant impacts on sentencing, are resource-intensive to develop and implement (in a context in which resources are highly limited), and merely pay lip service to addressing the crisis of mass incarceration. Grassroots organizations such as CADBI have been promoting low-tech liberatory policy changes for decades, 
	including abolishing cash bail, 
	releasing elderly populations from prison, and reinvesting money in schools and communities. Unlike risk assessment instruments, such measures do not rely on individual judges' alignment with policy goals and have robust empirical support for reducing prison populations.\footnote{See \cite{zhou_empirical_2021} and \cite{note_note_2018} for empirical discussions of bail reform.} 
	





	\section{Conclusion}
	
	In this paper, I presented a qualitative study of criminal court judges, probation officers, and Pennsylvania Sentencing Commission staff; the study's interview questions were designed with the assistance of the community organization CADBI. I found that judges ignored the tool, a result of the tool's lack of utility and shortcomings in how information is disseminated to judges, rather than a mere distrust of the tool or a fear of automation. This lack of utility, in turn, was the interplay of organizational factors and competing interests, which illustrates the importance of an organizational perspective on scholarship on algorithm aversion and resistance. This study adds to the empirical scholarship on risk assessment instruments' on-the-ground impacts and invites a departure from the speculative discourse around AI-centric criminal justice reforms. Evidently, algorithmic decision-making systems are not immune to the shortcomings of other bureaucratic changes. 
	
	%
	
	The Sentence Risk Assessment Instrument was the locus of considerable time and taxpayer dollars; it was in development for nearly a decade following a 2010 state legislative mandate for adopting a risk assessment tool for sentencing. Despite having no impact, the final version of the tool satisfies this mandate, producing the false impression that some evidence-based measure has been taken to address Pennsylvania's crisis of mass incarceration and racial disparities in sentencing. This study adds to an empirically-informed argument against reforms like these, which can help direct attention toward decarceration efforts that are less costly -- and actually work. 
	
	\begin{acks}
		Thank you to the Coalition to Abolish Death by Incarceration (CADBI) for their support throughout this project. I would like to thank Colin Allen, David Danks, Dorothea Anagnostopoulos, Camilo Ruiz, Sarah Brayne, Cierra Robson, Maria Ryabova, Sam Plummer, Rhys Hester, my anonymous reviewers, and members of my dissertation writing group for their invaluable feedback and advice on this project. This work was supported by the Horowitz Foundation for Social Policy and the University of Pittsburgh Year of Data and Society.
	\end{acks}

	
\bibliographystyle{ACM-Reference-Format}
\bibliography{Law+ML}


\begin{thebibliography}{49}


\ifx \showCODEN    \undefined \def \showCODEN     #1{\unskip}     \fi
\ifx \showDOI      \undefined \def \showDOI       #1{#1}\fi
\ifx \showISBNx    \undefined \def \showISBNx     #1{\unskip}     \fi
\ifx \showISBNxiii \undefined \def \showISBNxiii  #1{\unskip}     \fi
\ifx \showISSN     \undefined \def \showISSN      #1{\unskip}     \fi
\ifx \showLCCN     \undefined \def \showLCCN      #1{\unskip}     \fi
\ifx \shownote     \undefined \def \shownote      #1{#1}          \fi
\ifx \showarticletitle \undefined \def \showarticletitle #1{#1}   \fi
\ifx \showURL      \undefined \def \showURL       {\relax}        \fi
\providecommand\bibfield[2]{#2}
\providecommand\bibinfo[2]{#2}
\providecommand\natexlab[1]{#1}
\providecommand\showeprint[2][]{arXiv:#2}

\bibitem[{ACLU of Pennsylvania}(2019)]%
        {aclu_of_pennsylvania_testimony_2019}
\bibfield{author}{\bibinfo{person}{{ACLU of Pennsylvania}}.}
  \bibinfo{year}{2019}\natexlab{}.
\newblock \showarticletitle{Testimony for the {Pennsylvania} {Commission} on
  {Sentencing} {Regarding} the {July} 12, 2019, {Proposed} {Sentence} {Risk}
  {Assessment} {Tool}}.
\newblock \bibinfo{journal}{\emph{2019 08 Testimony (49 PaB 3718)}}
  (\bibinfo{year}{2019}).
\newblock
\newblock
\shownote{2019 08 Testimony (49 PaB 3718), Pennsylvania Commission on
  Sentencing}.


\bibitem[Akpinar et~al\mbox{.}(2021)]%
        {akpinar_effect_2021}
\bibfield{author}{\bibinfo{person}{Nil-Jana Akpinar}, \bibinfo{person}{Maria
  De-Arteaga}, {and} \bibinfo{person}{Alexandra Chouldechova}.}
  \bibinfo{year}{2021}\natexlab{}.
\newblock \showarticletitle{The effect of differential victim crime reporting
  on predictive policing systems}. In \bibinfo{booktitle}{\emph{Proceedings of
  the 2021 {ACM} {Conference} on {Fairness}, {Accountability}, and
  {Transparency}}} \emph{(\bibinfo{series}{{FAccT} '21})}.
  \bibinfo{publisher}{Association for Computing Machinery},
  \bibinfo{address}{New York, NY, USA}, \bibinfo{pages}{838--849}.
\newblock
\showISBNx{978-1-4503-8309-7}
\urldef\tempurl%
\url{https://doi.org/10.1145/3442188.3445877}
\showDOI{\tempurl}


\bibitem[Albright(2019)]%
        {albright_if_2019}
\bibfield{author}{\bibinfo{person}{Alex Albright}.}
  \bibinfo{year}{2019}\natexlab{}.
\newblock \bibinfo{title}{If {You} {Give} a {Judge} a {Risk} {Score}:
  {Evidence} from {Kentucky} {Bail} {Decisions}}.  (\bibinfo{year}{2019}).
\newblock


\bibitem[Alkhatib and Bernstein(2019)]%
        {alkhatib_street-level_2019}
\bibfield{author}{\bibinfo{person}{Ali Alkhatib} {and} \bibinfo{person}{Michael
  Bernstein}.} \bibinfo{year}{2019}\natexlab{}.
\newblock \showarticletitle{Street-{Level} {Algorithms}: {A} {Theory} at the
  {Gaps} {Between} {Policy} and {Decisions}}. In
  \bibinfo{booktitle}{\emph{Proceedings of the 2019 {CHI} {Conference} on
  {Human} {Factors} in {Computing} {Systems}}} \emph{(\bibinfo{series}{{CHI}
  '19})}. \bibinfo{publisher}{Association for Computing Machinery},
  \bibinfo{address}{New York, NY, USA}, \bibinfo{pages}{1--13}.
\newblock
\showISBNx{978-1-4503-5970-2}
\urldef\tempurl%
\url{https://doi.org/10.1145/3290605.3300760}
\showDOI{\tempurl}


\bibitem[Angwin et~al\mbox{.}(2016)]%
        {angwin_machine_2016}
\bibfield{author}{\bibinfo{person}{Julia Angwin}, \bibinfo{person}{Jeff
  Larson}, \bibinfo{person}{Surya Mattu}, {and} \bibinfo{person}{Lauren
  Kirchner}.} \bibinfo{year}{2016}\natexlab{}.
\newblock \showarticletitle{Machine {Bias}: {There}'s software used across the
  country to predict future criminals. {And} it's biased against blacks}.
\newblock \bibinfo{journal}{\emph{ProPublica}} (\bibinfo{year}{2016}).
\newblock
\newblock
\shownote{https://www.propublica.org/article/machine-bias-risk-assessments-in-criminal-sentencing}.


\bibitem[Becerril et~al\mbox{.}(2019)]%
        {becerril_validation_2019}
\bibfield{author}{\bibinfo{person}{David~Mitre Becerril},
  \bibinfo{person}{Chris Bell}, \bibinfo{person}{Katie LeFevre},
  \bibinfo{person}{Laurel Lin}, \bibinfo{person}{Wilson Mui},
  \bibinfo{person}{Karan Shah}, {and} \bibinfo{person}{Matt Hannigan}.}
  \bibinfo{year}{2019}\natexlab{}.
\newblock \showarticletitle{Validation and {Assessment} of {Pennsylvania}’s
  {Risk} {Assessment} {Instrument}, {Pennsylvania} {Commission} on
  {Sentencing}}.
\newblock \bibinfo{journal}{\emph{Heinz College System Synthesis Project}}
  (\bibinfo{date}{May} \bibinfo{year}{2019}).
\newblock


\bibitem[Brayne(2020)]%
        {brayne_predict_2020}
\bibfield{author}{\bibinfo{person}{Sarah Brayne}.}
  \bibinfo{year}{2020}\natexlab{}.
\newblock \bibinfo{booktitle}{\emph{Predict and {Surveil}: {Data},
  {Discretion}, and the {Future} of {Policing}}}.
\newblock \bibinfo{publisher}{Oxford University Press}.
\newblock
\showISBNx{978-0-19-068409-9}


\bibitem[Brayne and Christin(2020)]%
        {brayne_technologies_2020}
\bibfield{author}{\bibinfo{person}{Sarah Brayne} {and} \bibinfo{person}{Angèle
  Christin}.} \bibinfo{year}{2020}\natexlab{}.
\newblock \showarticletitle{Technologies of {Crime} {Prediction}: {The}
  {Reception} of {Algorithms} in {Policing} and {Criminal} {Courts}}.
\newblock \bibinfo{journal}{\emph{Social Problems}} (\bibinfo{year}{2020}).
\newblock
\urldef\tempurl%
\url{https://doi.org/10.1093/socpro/spaa004}
\showDOI{\tempurl}


\bibitem[Brown et~al\mbox{.}(2019)]%
        {brown_toward_2019}
\bibfield{author}{\bibinfo{person}{Anna Brown}, \bibinfo{person}{Alexandra
  Chouldechova}, \bibinfo{person}{Emily Putnam-Hornstein},
  \bibinfo{person}{Andrew Tobin}, {and} \bibinfo{person}{Rhema Vaithianathan}.}
  \bibinfo{year}{2019}\natexlab{}.
\newblock \showarticletitle{Toward {Algorithmic} {Accountability} in {Public}
  {Services}: {A} {Qualitative} {Study} of {Affected} {Community}
  {Perspectives} on {Algorithmic} {Decision}-making in {Child} {Welfare}
  {Services}}. In \bibinfo{booktitle}{\emph{Proceedings of the 2019 {CHI}
  {Conference} on {Human} {Factors} in {Computing} {Systems}}}
  \emph{(\bibinfo{series}{{CHI} '19})}. \bibinfo{publisher}{Association for
  Computing Machinery}, \bibinfo{address}{New York, NY, USA},
  \bibinfo{pages}{1--12}.
\newblock
\showISBNx{978-1-4503-5970-2}
\urldef\tempurl%
\url{https://doi.org/10.1145/3290605.3300271}
\showDOI{\tempurl}


\bibitem[Cheng et~al\mbox{.}(2022)]%
        {cheng_how_2022}
\bibfield{author}{\bibinfo{person}{Hao-Fei Cheng}, \bibinfo{person}{Logan
  Stapleton}, \bibinfo{person}{Anna Kawakami}, \bibinfo{person}{Venkatesh
  Sivaraman}, \bibinfo{person}{Yanghuidi Cheng}, \bibinfo{person}{Diana Qing},
  \bibinfo{person}{Adam Perer}, \bibinfo{person}{Kenneth Holstein},
  \bibinfo{person}{Zhiwei~Steven Wu}, {and} \bibinfo{person}{Haiyi Zhu}.}
  \bibinfo{year}{2022}\natexlab{}.
\newblock \showarticletitle{How {Child} {Welfare} {Workers} {Reduce} {Racial}
  {Disparities} in {Algorithmic} {Decisions}}. In
  \bibinfo{booktitle}{\emph{Proceedings of the 2022 {CHI} {Conference} on
  {Human} {Factors} in {Computing} {Systems}}} \emph{(\bibinfo{series}{{CHI}
  '22})}. \bibinfo{publisher}{Association for Computing Machinery},
  \bibinfo{address}{New York, NY, USA}, \bibinfo{pages}{1--22}.
\newblock
\showISBNx{978-1-4503-9157-3}
\urldef\tempurl%
\url{https://doi.org/10.1145/3491102.3501831}
\showDOI{\tempurl}


\bibitem[Chouldechova(2017)]%
        {chouldechova_fair_2017}
\bibfield{author}{\bibinfo{person}{Alexandra Chouldechova}.}
  \bibinfo{year}{2017}\natexlab{}.
\newblock \showarticletitle{Fair {Prediction} with {Disparate} {Impact}: {A}
  {Study} of {Bias} in {Recidivism} {Prediction} {Instruments}}.
\newblock \bibinfo{journal}{\emph{Big Data}} \bibinfo{volume}{5},
  \bibinfo{number}{2} (\bibinfo{year}{2017}), \bibinfo{pages}{153--163}.
\newblock
\showISSN{2167-6461}
\urldef\tempurl%
\url{https://doi.org/10.1089/big.2016.0047}
\showDOI{\tempurl}


\bibitem[Christin(2017)]%
        {christin_algorithms_2017}
\bibfield{author}{\bibinfo{person}{Angèle Christin}.}
  \bibinfo{year}{2017}\natexlab{}.
\newblock \showarticletitle{Algorithms in practice: {Comparing} web journalism
  and criminal justice}.
\newblock \bibinfo{journal}{\emph{Big Data \& Society}} \bibinfo{volume}{4},
  \bibinfo{number}{2} (\bibinfo{date}{Dec.} \bibinfo{year}{2017}),
  \bibinfo{pages}{2053951717718855}.
\newblock
\showISSN{2053-9517}
\urldef\tempurl%
\url{https://doi.org/10.1177/2053951717718855}
\showDOI{\tempurl}
\newblock
\shownote{Publisher: SAGE Publications Ltd}.


\bibitem[{Coalition to Abolish Death by Incarceration}(2019)]%
        {coalition_to_abolish_death_by_incarceration_proposed_2019}
\bibfield{author}{\bibinfo{person}{{Coalition to Abolish Death by
  Incarceration}}.} \bibinfo{year}{2019}\natexlab{}.
\newblock \showarticletitle{Proposed {Risk} {Assessment} {Instrument} {Public}
  {Hearing}}.
\newblock \bibinfo{journal}{\emph{08 Testimony (49 PaB 3718)}}
  (\bibinfo{year}{2019}).
\newblock
\newblock
\shownote{Quizz Cozzens, in testimony to the Pennsylvania Commission on
  Sentencing,
  https://pcs.la.psu.edu/guidelines-statutes/risk-assessment/sentence-risk-assessment-proposals-and-testimony/}.


\bibitem[{County Chief Adult Probation and Parole Officers Association of
  Pennsylvania}(2019)]%
  {county_chief_adult_probation_and_parole_officers_association_of_pennsylvania_re_2019}
\bibfield{author}{\bibinfo{person}{{County Chief Adult Probation and Parole
  Officers Association of Pennsylvania}}.} \bibinfo{year}{2019}\natexlab{}.
\newblock \showarticletitle{Re: {Proposed} {Sentence} {Risk} {Assessment}
  {Instrument}}.
\newblock \bibinfo{journal}{\emph{2019 08 Testimony (49 PaB 3718)}}
  (\bibinfo{year}{2019}).
\newblock
\newblock
\shownote{2019 08 Testimony (49 PaB 3718), Pennsylvania Commission on
  Sentencing}.


\bibitem[De-Arteaga et~al\mbox{.}(2020)]%
        {de-arteaga_case_2020}
\bibfield{author}{\bibinfo{person}{Maria De-Arteaga}, \bibinfo{person}{Riccardo
  Fogliato}, {and} \bibinfo{person}{Alexandra Chouldechova}.}
  \bibinfo{year}{2020}\natexlab{}.
\newblock \showarticletitle{A {Case} for {Humans}-in-the-{Loop}: {Decisions} in
  the {Presence} of {Erroneous} {Algorithmic} {Scores}}. In
  \bibinfo{booktitle}{\emph{Proceedings of the 2020 {CHI} {Conference} on
  {Human} {Factors} in {Computing} {Systems}}} \emph{(\bibinfo{series}{{CHI}
  '20})}. \bibinfo{publisher}{Association for Computing Machinery},
  \bibinfo{address}{New York, NY, USA}, \bibinfo{pages}{1--12}.
\newblock
\showISBNx{978-1-4503-6708-0}
\urldef\tempurl%
\url{https://doi.org/10.1145/3313831.3376638}
\showDOI{\tempurl}


\bibitem[Dietvorst et~al\mbox{.}(2015)]%
        {dietvorst_algorithm_2015}
\bibfield{author}{\bibinfo{person}{Berkeley~J. Dietvorst},
  \bibinfo{person}{Joseph~P. Simmons}, {and} \bibinfo{person}{Cade Massey}.}
  \bibinfo{year}{2015}\natexlab{}.
\newblock \showarticletitle{Algorithm aversion: {People} erroneously avoid
  algorithms after seeing them err}.
\newblock \bibinfo{journal}{\emph{Journal of Experimental Psychology: General}}
   \bibinfo{volume}{144} (\bibinfo{year}{2015}), \bibinfo{pages}{114--126}.
\newblock
\showISSN{1939-2222}
\urldef\tempurl%
\url{https://doi.org/10.1037/xge0000033}
\showDOI{\tempurl}
\newblock
\shownote{Place: US Publisher: American Psychological Association}.


\bibitem[Espeland and Vannebo(2007)]%
        {espeland_accountability_2007}
\bibfield{author}{\bibinfo{person}{Wendy~Nelson Espeland} {and}
  \bibinfo{person}{Berit~Irene Vannebo}.} \bibinfo{year}{2007}\natexlab{}.
\newblock \showarticletitle{Accountability, quantification, and law}.
\newblock In \bibinfo{booktitle}{\emph{Annual {Review} of {Law} and {Social}
  {Science}}}, \bibfield{editor}{\bibinfo{person}{John Hagan},
  \bibinfo{person}{Kim~Lane Scheppele}, {and} \bibinfo{person}{Tom Tyler}}
  (Eds.). \bibinfo{pages}{21--43}.
\newblock
\showISBNx{978-0-8243-4103-9}
\urldef\tempurl%
\url{https://doi.org/10.1146/annurev.lawsocsci.2.081805.105908}
\showDOI{\tempurl}


\bibitem[Fogliato et~al\mbox{.}(2021)]%
        {fogliato_impact_2021}
\bibfield{author}{\bibinfo{person}{Riccardo Fogliato},
  \bibinfo{person}{Alexandra Chouldechova}, {and} \bibinfo{person}{Zachary
  Lipton}.} \bibinfo{year}{2021}\natexlab{}.
\newblock \showarticletitle{The {Impact} of {Algorithmic} {Risk} {Assessments}
  on {Human} {Predictions} and its {Analysis} via {Crowdsourcing} {Studies}}.
\newblock \bibinfo{journal}{\emph{Proceedings of the ACM on Human-Computer
  Interaction}} \bibinfo{volume}{5}, \bibinfo{number}{CSCW2}
  (\bibinfo{date}{Oct.} \bibinfo{year}{2021}), \bibinfo{pages}{428:1--428:24}.
\newblock
\urldef\tempurl%
\url{https://doi.org/10.1145/3479572}
\showDOI{\tempurl}


\bibitem[Galison(2019)]%
        {galison_algorists_2019}
\bibfield{author}{\bibinfo{person}{Peter~L. Galison}.}
  \bibinfo{year}{2019}\natexlab{}.
\newblock \showarticletitle{Algorists {Dream} of {Objectivity}}.
\newblock In \bibinfo{booktitle}{\emph{Possible {Minds}: 25 {Ways} of {Looking}
  at {AI}}}, \bibfield{editor}{\bibinfo{person}{John Brockman}} (Ed.).
  \bibinfo{publisher}{Penguin Publishing Group}.
\newblock


\bibitem[Garrett and Monahan(2020)]%
        {garrett_judging_2020}
\bibfield{author}{\bibinfo{person}{Brandon Garrett} {and} \bibinfo{person}{John
  Monahan}.} \bibinfo{year}{2020}\natexlab{}.
\newblock \showarticletitle{Judging {Risk}}.
\newblock \bibinfo{journal}{\emph{California Law Review}}
  \bibinfo{volume}{108}, \bibinfo{number}{2} (\bibinfo{date}{Jan.}
  \bibinfo{year}{2020}), \bibinfo{pages}{439--493}.
\newblock
\urldef\tempurl%
\url{https://scholarship.law.duke.edu/faculty_scholarship/3882}
\showURL{%
\tempurl}


\bibitem[Glaser et~al\mbox{.}(2021)]%
        {glaser_biography_2021}
\bibfield{author}{\bibinfo{person}{Vern~L. Glaser}, \bibinfo{person}{Neil
  Pollock}, {and} \bibinfo{person}{Luciana D’Adderio}.}
  \bibinfo{year}{2021}\natexlab{}.
\newblock \showarticletitle{The {Biography} of an {Algorithm}: {Performing}
  algorithmic technologies in organizations}.
\newblock \bibinfo{journal}{\emph{Organization Theory}} \bibinfo{volume}{2},
  \bibinfo{number}{2} (\bibinfo{date}{April} \bibinfo{year}{2021}),
  \bibinfo{pages}{26317877211004609}.
\newblock
\showISSN{2631-7877}
\urldef\tempurl%
\url{https://doi.org/10.1177/26317877211004609}
\showDOI{\tempurl}
\newblock
\shownote{Publisher: SAGE Publications Ltd}.


\bibitem[Gray and Suri(2019)]%
        {gray_ghost_2019}
\bibfield{author}{\bibinfo{person}{Mary~L. Gray} {and}
  \bibinfo{person}{Siddharth Suri}.} \bibinfo{year}{2019}\natexlab{}.
\newblock \bibinfo{booktitle}{\emph{Ghost {Work}: {How} to {Stop} {Silicon}
  {Valley} from {Building} a {New} {Global} {Underclass}}}.
\newblock \bibinfo{publisher}{HarperCollins}.
\newblock
\showISBNx{978-1-328-56628-7}


\bibitem[Gross(2019)]%
        {gross_pennsylvanias_2019}
\bibfield{author}{\bibinfo{person}{Paige Gross}.}
  \bibinfo{year}{2019}\natexlab{}.
\newblock \bibinfo{title}{Pennsylvania's controversial risk-assessment tool was
  just approved}.
\newblock
\newblock
\urldef\tempurl%
\url{https://technical.ly/civic-news/pennsylvanias-controversial-sentencing-risk-assessment-tool-was-just-approved/}
\showURL{%
\tempurl}


\bibitem[Hannah-Moffat(2013)]%
        {hannah-moffat_actuarial_2013}
\bibfield{author}{\bibinfo{person}{Kelly Hannah-Moffat}.}
  \bibinfo{year}{2013}\natexlab{}.
\newblock \showarticletitle{Actuarial {Sentencing}: {An} “{Unsettled}”
  {Proposition}}.
\newblock \bibinfo{journal}{\emph{Justice Quarterly}} \bibinfo{volume}{30},
  \bibinfo{number}{2} (\bibinfo{date}{April} \bibinfo{year}{2013}),
  \bibinfo{pages}{270--296}.
\newblock
\showISSN{0741-8825}
\urldef\tempurl%
\url{https://doi.org/10.1080/07418825.2012.682603}
\showDOI{\tempurl}


\bibitem[Harcourt(2008)]%
        {harcourt_against_2008}
\bibfield{author}{\bibinfo{person}{Bernard~E. Harcourt}.}
  \bibinfo{year}{2008}\natexlab{}.
\newblock \bibinfo{booktitle}{\emph{Against {Prediction}: {Profiling},
  {Policing}, and {Punishing} in an {Actuarial} {Age}}}.
\newblock \bibinfo{publisher}{University of Chicago Press}.
\newblock
\showISBNx{978-0-226-31599-7}


\bibitem[Harding(1992)]%
        {harding_rethinking_1992}
\bibfield{author}{\bibinfo{person}{Sandra Harding}.}
  \bibinfo{year}{1992}\natexlab{}.
\newblock \showarticletitle{Rethinking {Standpoint} {Epistemology}: {What} is
  ``{Strong} {Objectivity}?"}.
\newblock \bibinfo{journal}{\emph{The Centennial Review}} \bibinfo{volume}{36},
  \bibinfo{number}{3} (\bibinfo{year}{1992}), \bibinfo{pages}{437--470}.
\newblock
\showISSN{0162-0177}
\urldef\tempurl%
\url{https://www.jstor.org/stable/23739232}
\showURL{%
\tempurl}


\bibitem[{Human Rights Watch}(2017)]%
        {human_rights_watch_not_2017}
\bibfield{author}{\bibinfo{person}{{Human Rights Watch}}.}
  \bibinfo{year}{2017}\natexlab{}.
\newblock \bibinfo{booktitle}{\emph{“{Not} in it for {Justice}”: {How}
  {California}’s {Pretrial} {Detention} and {Bail} {System} {Unfairly}
  {Punishes} {Poor} {People}}}.
\newblock \bibinfo{type}{{T}echnical {R}eport}.
\newblock
\urldef\tempurl%
\url{https://www.hrw.org/report/2017/04/11/not-it-justice/how-californias-pretrial-detention-and-bail-system-unfairly}
\showURL{%
\tempurl}


\bibitem[Levy et~al\mbox{.}(2021)]%
        {levy_algorithms_2021}
\bibfield{author}{\bibinfo{person}{Karen Levy}, \bibinfo{person}{Kyla~E.
  Chasalow}, {and} \bibinfo{person}{Sarah Riley}.}
  \bibinfo{year}{2021}\natexlab{}.
\newblock \showarticletitle{Algorithms and {Decision}-{Making} in the {Public}
  {Sector}}.
\newblock \bibinfo{journal}{\emph{Annual Review of Law and Social Science}}
  \bibinfo{volume}{17}, \bibinfo{number}{1} (\bibinfo{year}{2021}),
  \bibinfo{pages}{309--334}.
\newblock
\urldef\tempurl%
\url{https://doi.org/10.1146/annurev-lawsocsci-041221-023808}
\showDOI{\tempurl}


\bibitem[Lipsky(1980)]%
        {lipsky_street-level_1980}
\bibfield{author}{\bibinfo{person}{Michael Lipsky}.}
  \bibinfo{year}{1980}\natexlab{}.
\newblock \bibinfo{booktitle}{\emph{Street-{Level} {Bureaucracy}: {The}
  {Dilemmas} of the {Individual} in {Public} {Service}}}.
\newblock \bibinfo{publisher}{Russell Sage Foundation}.
\newblock
\showISBNx{978-0-87154-526-8}


\bibitem[Mahmud et~al\mbox{.}(2022)]%
        {mahmud_what_2022}
\bibfield{author}{\bibinfo{person}{Hasan Mahmud}, \bibinfo{person}{A.~K.
  M.~Najmul Islam}, \bibinfo{person}{Syed~Ishtiaque Ahmed}, {and}
  \bibinfo{person}{Kari Smolander}.} \bibinfo{year}{2022}\natexlab{}.
\newblock \showarticletitle{What influences algorithmic decision-making? {A}
  systematic literature review on algorithm aversion}.
\newblock \bibinfo{journal}{\emph{Technological Forecasting and Social Change}}
   \bibinfo{volume}{175} (\bibinfo{date}{Feb.} \bibinfo{year}{2022}),
  \bibinfo{pages}{121390}.
\newblock
\showISSN{0040-1625}
\urldef\tempurl%
\url{https://doi.org/10.1016/j.techfore.2021.121390}
\showDOI{\tempurl}


\bibitem[Merriam and Tisdell(2015)]%
        {merriam_qualitative_2015}
\bibfield{author}{\bibinfo{person}{Sharan~B. Merriam} {and}
  \bibinfo{person}{Elizabeth~J. Tisdell}.} \bibinfo{year}{2015}\natexlab{}.
\newblock \bibinfo{booktitle}{\emph{Qualitative {Research}: {A} {Guide} to
  {Design} and {Implementation}}}.
\newblock \bibinfo{publisher}{John Wiley \& Sons}.
\newblock
\showISBNx{978-1-119-00361-8}


\bibitem[Meyer et~al\mbox{.}(2022)]%
        {meyer_flipping_2022}
\bibfield{author}{\bibinfo{person}{Mikaela Meyer}, \bibinfo{person}{Aaron
  Horowitz}, \bibinfo{person}{Erica Marshall}, {and} \bibinfo{person}{Kristian
  Lum}.} \bibinfo{year}{2022}\natexlab{}.
\newblock \showarticletitle{Flipping the {Script} on {Criminal} {Justice}
  {Risk} {Assessment}: {An} actuarial model for assessing the risk the federal
  sentencing system poses to defendants}. In \bibinfo{booktitle}{\emph{2022
  {ACM} {Conference} on {Fairness}, {Accountability}, and {Transparency}}}
  \emph{(\bibinfo{series}{{FAccT} '22})}. \bibinfo{publisher}{Association for
  Computing Machinery}, \bibinfo{address}{New York, NY, USA},
  \bibinfo{pages}{366--378}.
\newblock
\showISBNx{978-1-4503-9352-2}
\urldef\tempurl%
\url{https://doi.org/10.1145/3531146.3533104}
\showDOI{\tempurl}


\bibitem[Miles et~al\mbox{.}(2014)]%
        {miles_qualitative_2014}
\bibfield{author}{\bibinfo{person}{Matthew~B. Miles},
  \bibinfo{person}{A.~Michael Huberman}, {and} \bibinfo{person}{Johnny
  Saldana}.} \bibinfo{year}{2014}\natexlab{}.
\newblock \bibinfo{booktitle}{\emph{Qualitative {Data} {Analysis}}}.
\newblock \bibinfo{publisher}{SAGE}.
\newblock
\showISBNx{978-1-4522-5787-7}


\bibitem[Mittelstadt et~al\mbox{.}(2016)]%
        {mittelstadt_ethics_2016}
\bibfield{author}{\bibinfo{person}{Brent~Daniel Mittelstadt},
  \bibinfo{person}{Patrick Allo}, \bibinfo{person}{Mariarosaria Taddeo},
  \bibinfo{person}{Sandra Wachter}, {and} \bibinfo{person}{Luciano Floridi}.}
  \bibinfo{year}{2016}\natexlab{}.
\newblock \showarticletitle{The ethics of algorithms: {Mapping} the debate}.
\newblock \bibinfo{journal}{\emph{Big Data \& Society}} \bibinfo{volume}{3},
  \bibinfo{number}{2} (\bibinfo{date}{Dec.} \bibinfo{year}{2016}),
  \bibinfo{pages}{2053951716679679}.
\newblock
\showISSN{2053-9517}
\urldef\tempurl%
\url{https://doi.org/10.1177/2053951716679679}
\showDOI{\tempurl}
\newblock
\shownote{Publisher: SAGE Publications Ltd}.


\bibitem[{Note}(2018)]%
        {note_note_2018}
\bibfield{author}{\bibinfo{person}{{Note}}.} \bibinfo{year}{2018}\natexlab{}.
\newblock \showarticletitle{Note, {Bail} {Reform} and {Risk} {Assessment}:
  {The} {Cautionary} {Tale} of {Federal} {Sentencing}}.
\newblock \bibinfo{journal}{\emph{Harvard Law Review}} \bibinfo{volume}{131},
  \bibinfo{number}{1125} (\bibinfo{year}{2018}).
\newblock


\bibitem[Parker et~al\mbox{.}(2019)]%
        {parker_snowball_2019}
\bibfield{author}{\bibinfo{person}{C. Parker}, \bibinfo{person}{S. Scott},
  {and} \bibinfo{person}{A. Geddes}.} \bibinfo{year}{2019}\natexlab{}.
\newblock \showarticletitle{Snowball {Sampling}}.
\newblock \bibinfo{journal}{\emph{SAGE Research Methods Foundations}}
  (\bibinfo{date}{Sept.} \bibinfo{year}{2019}).
\newblock
\urldef\tempurl%
\url{http://methods.sagepub.com/foundations/snowball-sampling}
\showURL{%
\tempurl}
\newblock
\shownote{Publisher: SAGE}.


\bibitem[{Pennsylvania Commission on Sentencing}(2019)]%
        {pennsylvania_commission_on_sentencing_adopted_2019}
\bibfield{author}{\bibinfo{person}{{Pennsylvania Commission on Sentencing}}.}
  \bibinfo{year}{2019}\natexlab{}.
\newblock \showarticletitle{Adopted {Sentence} {Risk} {Assessment}
  {Instrument}}.
\newblock \bibinfo{journal}{\emph{Title 204, Part VII, Chapter 305}}
  (\bibinfo{year}{2019}).
\newblock


\bibitem[{Pennsylvania Commission on Sentencing}(2021)]%
        {pennsylvania_commission_on_sentencing_commission_2021}
\bibfield{author}{\bibinfo{person}{{Pennsylvania Commission on Sentencing}}.}
  \bibinfo{year}{2021}\natexlab{}.
\newblock \bibinfo{title}{Commission {Policy} {Meeting}, {Annual} {Planning}
  {Meeting} {Slides} ({Part} 2, {Sentencing} {Risk} {Assessment} {Instrument}
  {Initial} {Analysis})}.
\newblock
\newblock
\newblock
\shownote{https://pcs.la.psu.edu/policy-administration/previous-commission-policy-meetings/}.


\bibitem[{Pennsylvania Commission on Sentencing}(2022)]%
        {pennsylvania_commission_on_sentencing_pennsylvania_2022}
\bibfield{author}{\bibinfo{person}{{Pennsylvania Commission on Sentencing}}.}
  \bibinfo{year}{2022}\natexlab{}.
\newblock \bibinfo{title}{Pennsylvania {Commission} on {Sentencing} {Website},
  https://pcs.la.psu.edu/}.
\newblock
\newblock


\bibitem[Porter(1995)]%
        {porter_trust_1995}
\bibfield{author}{\bibinfo{person}{Theodore~M. Porter}.}
  \bibinfo{year}{1995}\natexlab{}.
\newblock \bibinfo{booktitle}{\emph{Trust in numbers: the pursuit of
  objectivity in science and public life}}.
\newblock \bibinfo{publisher}{Princeton University Press},
  \bibinfo{address}{Princeton, N.J}.
\newblock
\showISBNx{978-0-691-03776-9}


\bibitem[Pruss(2021)]%
        {pruss_mechanical_2021-1}
\bibfield{author}{\bibinfo{person}{Dasha Pruss}.}
  \bibinfo{year}{2021}\natexlab{}.
\newblock \showarticletitle{Mechanical {Jurisprudence} and {Domain}
  {Distortion}: {How} {Predictive} {Algorithms} {Warp} the {Law}}.
\newblock \bibinfo{journal}{\emph{Philosophy of Science}} \bibinfo{volume}{88},
  \bibinfo{number}{5} (\bibinfo{date}{Dec.} \bibinfo{year}{2021}),
  \bibinfo{pages}{1101--1112}.
\newblock
\showISSN{0031-8248, 1539-767X}
\urldef\tempurl%
\url{https://doi.org/10.1086/715512}
\showDOI{\tempurl}
\newblock
\shownote{Publisher: Cambridge University Press}.


\bibitem[Sassaman(2018)]%
        {sassaman_testimony_2018}
\bibfield{author}{\bibinfo{person}{Hannah Sassaman}.}
  \bibinfo{year}{2018}\natexlab{}.
\newblock \showarticletitle{Testimony}.
\newblock \bibinfo{journal}{\emph{06 Testimony (48 PaB 2367)}}
  (\bibinfo{year}{2018}).
\newblock
\newblock
\shownote{Pennsylvania Sentencing Commission,
  https://pcs.la.psu.edu/guidelines-statutes/risk-assessment/sentence-risk-assessment-proposals-and-testimony/}.


\bibitem[Sassaman(2019)]%
        {sassaman_pennsylvanias_2019}
\bibfield{author}{\bibinfo{person}{Hannah Sassaman}.}
  \bibinfo{year}{2019}\natexlab{}.
\newblock \showarticletitle{Pennsylvania's proposed risk-assessment algorithm
  is racist}.
\newblock \bibinfo{journal}{\emph{The Inquirer}} (\bibinfo{date}{Sept.}
  \bibinfo{year}{2019}).
\newblock
\newblock
\shownote{https://www.inquirer.com/opinion/commentary/pennsylvania-sentencing-commission-rat-risk-assessment-20190904.html}.


\bibitem[Sloan et~al\mbox{.}(2018)]%
        {sloan_effect_2018}
\bibfield{author}{\bibinfo{person}{CarlyWill Sloan}, \bibinfo{person}{George
  Naufal}, {and} \bibinfo{person}{Heather Caspers}.}
  \bibinfo{year}{2018}\natexlab{}.
\newblock \bibinfo{title}{The {Effect} of {Risk} {Assessment} {Scores} on
  {Judicial} {Behavior} and {Defendant} {Outcomes}}.
\newblock
\newblock
\urldef\tempurl%
\url{https://doi.org/10.2139/ssrn.3301699}
\showDOI{\tempurl}


\bibitem[Small(2009)]%
        {small_how_2009}
\bibfield{author}{\bibinfo{person}{Mario~Luis Small}.}
  \bibinfo{year}{2009}\natexlab{}.
\newblock \showarticletitle{`{How} many cases do {I} need?': {On} science and
  the logic of case selection in field-based research}.
\newblock \bibinfo{journal}{\emph{Ethnography}} \bibinfo{volume}{10},
  \bibinfo{number}{1} (\bibinfo{date}{March} \bibinfo{year}{2009}),
  \bibinfo{pages}{5--38}.
\newblock
\showISSN{1466-1381}
\urldef\tempurl%
\url{https://doi.org/10.1177/1466138108099586}
\showDOI{\tempurl}
\newblock
\shownote{Publisher: SAGE Publications}.


\bibitem[Stevenson(2018)]%
        {stevenson_assessing_2018}
\bibfield{author}{\bibinfo{person}{Megan~T. Stevenson}.}
  \bibinfo{year}{2018}\natexlab{}.
\newblock \showarticletitle{Assessing {Risk} {Assessment} in {Action}}.
\newblock \bibinfo{journal}{\emph{Minnesota Law Review}}  \bibinfo{volume}{103}
  (\bibinfo{year}{2018}), \bibinfo{pages}{303}.
\newblock
\urldef\tempurl%
\url{https://papers.ssrn.com/abstract=3016088}
\showURL{%
\tempurl}


\bibitem[Stevenson and Doleac(2021)]%
        {stevenson_algorithmic_2021}
\bibfield{author}{\bibinfo{person}{Megan~T. Stevenson} {and}
  \bibinfo{person}{Jennifer~L. Doleac}.} \bibinfo{year}{2021}\natexlab{}.
\newblock \bibinfo{booktitle}{\emph{Algorithmic {Risk} {Assessment} in the
  {Hands} of {Humans}}}.
\newblock \bibinfo{type}{{SSRN} {Scholarly} {Paper}} ID 3489440.
  \bibinfo{institution}{Social Science Research Network},
  \bibinfo{address}{Rochester, NY}.
\newblock
\urldef\tempurl%
\url{https://papers.ssrn.com/abstract=3489440}
\showURL{%
\tempurl}


\bibitem[Weiss(1995)]%
        {weiss_learning_1995}
\bibfield{author}{\bibinfo{person}{Robert~S. Weiss}.}
  \bibinfo{year}{1995}\natexlab{}.
\newblock \bibinfo{booktitle}{\emph{Learning {From} {Strangers}: {The} {Art}
  and {Method} of {Qualitative} {Interview} {Studies}}}.
\newblock \bibinfo{publisher}{Simon and Schuster}.
\newblock
\showISBNx{978-1-4391-0698-3}


\bibitem[Zhou et~al\mbox{.}(2021)]%
        {zhou_empirical_2021}
\bibfield{author}{\bibinfo{person}{Angela Zhou}, \bibinfo{person}{Andrew Koo},
  \bibinfo{person}{Nathan Kallus}, \bibinfo{person}{Rene Ropac},
  \bibinfo{person}{Richard Peterson}, \bibinfo{person}{Stephen Koppel}, {and}
  \bibinfo{person}{Tiffany Bergin}.} \bibinfo{year}{2021}\natexlab{}.
\newblock \bibinfo{title}{An {Empirical} {Evaluation} of the {Impact} of {New}
  {York}'s {Bail} {Reform} on {Crime} {Using} {Synthetic} {Controls}}.
\newblock
\newblock
\urldef\tempurl%
\url{https://doi.org/10.2139/ssrn.3964067}
\showDOI{\tempurl}


\end{thebibliography}

\appendix

\section{Demographic Survey}
\label{sec:survey}

\begin{enumerate}
	\item How long have you been a judge? \rule{2cm}{0.4pt}
	\item Please indicate your gender.
	\begin{enumerate}
		\item Female
		\item Male
	\end{enumerate}
	\item Please indicate your age.
	\begin{enumerate}
		\item 18--29
		\item 30--39
		\item 40--49
		\item 50--59
		\item 60--69
		\item 70-79
		\item 80+
	\end{enumerate}
	\item Which of the following best describes you?
	\begin{enumerate}
		\item Asian or Pacific Islander
		\item Black or African American
		\item Hispanic or Latino
		\item Native American or Alaskan Native
		\item White or Caucasian
		\item Other: \rule{2cm}{0.4pt}
	\end{enumerate}
	\item Do you think of yourself as a Republican, a Democrat, an Independent, or something else?
	\begin{enumerate}
		\item Republican
		\item Democrat
		\item Independent
		\item Other: \rule{2cm}{0.4pt}
	\end{enumerate}

\end{enumerate}

\section{Interview Guide for Judges}
\label{sec:interview_guide}


\noindent Thanks so much for taking the time to talk to me today. As I mentioned in the letter, I have been talking to judges statewide about how they are using the Pennsylvania Sentencing Commission's new Sentence Risk Assessment Instrument. The purpose of the study is to understand the impacts the tool has had on judicial practice. \vspace{.6em}

\noindent I'm going to ask you some open-ended questions about your professional background, your sentencing process, and your experience with this specific tool. How does that sound to you? \vspace{.6em}

\noindent [Offer to answer questions about the study, ask permission to record audio from the meeting, then start recording. Provide the following information on the tape:]

\begin{itemize}
	\item[--] court site/location
	\item[--] judge name 
	\item[--] date 
	\item[--] interview number 
\end{itemize}

\noindent \begin{enumerate}
	\item \textbf{Professional background} [3 minutes]
	
	I'd like to begin by hearing a bit about your professional background. Could you please tell me how you became a judge? [Keep this as brief as possible]
	
	Probes:
	\begin{itemize}
		\item Tell me about your prior work experiences related to being a judge.
		\item How long have you been a judge?
		\item How long in criminal?
	\end{itemize}

	\item \textbf{Sentencing process} [5-10 minutes] 
	
	Before we talk about the risk assessment tool, I'd like to hear about the process you typically go through when deciding a sentence.

	Probes:
	
	\begin{itemize}
		\item Factors you consider most important for deciding a sentence. Ask for a specific example: ``Can you tell me about a case from this past week?''
		
		\begin{itemize}
			\item Demographic factors? (e.g., age, past criminal history)
			\item Recidivism risk: considered/not considered, important/not important?
			\item (If recidivism considered) Most important factors for assessing recidivism risk?
		\end{itemize}


		\item When do you order a pre-sentence investigation report?
		
		\begin{itemize}
			\item What information do you receive in the PSI?
		\end{itemize}

	\end{itemize}

	\item \textbf{Instrument implementation/training} [10 minutes]
	
	Let's talk specifically about the Sentence Risk Assessment Instrument. Can you tell me how it was first introduced to you?
	
	Probes:
	\begin{itemize}
		\item What were your first impressions of the tool?
		\item Tell me about any training you received in using the tool.
		\item Did you have any concerns about the tool's introduction?
		\item Do you recall talking with your colleagues about the tool?
		\item When was the first time you saw a case where the tool applied?
	\end{itemize}

	\item \textbf{Instrument use} [5-10 minutes]
	
	I'd like to turn now to your actual experiences using the instrument. Could you walk me through what happens when you receive the tool's recommendations?
	
	Probes:
	
	\begin{itemize}
		\item What other information do you get at sentencing time?
		\item Where is the Sentence Risk Assessment information presented to you?
		\item How many cases have you seen so far? 
		\item Is the tool's recommendation something you typically make note of?
		\item What happens when the tool recommends seeking ``Additional Information''? (Ask for specific examples)
		
	\end{itemize}

	\item \textbf{Examples of changes (or lack thereof)} [5 minutes]
	
	I'm interested in hearing whether you've noticed any changes in your day-to-day work since the introduction of the tool. It would be helpful to hear specific examples of things the tool has and has not affected.
	
	\begin{itemize}
		\item When you see the ``Additional Information'' label, do you infer the defendant's risk level from this?
		
		\begin{itemize}
			\item (If yes) do you think this inference about recidivism risk level affects how you think about a case?
		\end{itemize}
		\item Can you give an example of a case with the ``Additional information'' label where you chose to order a pre-sentence investigation report? 
		\item Can you give an example of an ``Additional information'' case where you did not order a pre-sentence investigation report? 
		\item Can you give an example where using the tool changed the sentence you assigned?

		\begin{itemize}
			\item What was it about this additional information that changed your sentence?
		\end{itemize}

		\item Can you give an example of when using the tool had no effect on the sentence you assigned?
	\end{itemize}

	\item \textbf{Risk assessment in general} [5-10 minutes]
	
	I'd like to hear what you think about risk assessment tools in general. 
	
	\begin{itemize}
		\item (If judge mentions racial bias/disparities) Do you think this risk assessment tool could help with the disparities/make them worse?
		\item Do you feel that this tool helps judges identify appropriate candidates for alternative sentencing? Why or why not?
		\item In general, have you found the tool useful?
	\end{itemize}

	%


	\item \textbf{Thank you and conclude} [3 minutes]
	
	Thank you so much for taking the time to talk to me. This was very helpful.
	
	\begin{itemize}
		\item Is there anything you'd like to add that I haven't asked that you think is relevant to this project?
		\item Follow-up: After an interview I always find that I've forgotten to ask something. Would it be all right with you if I send you a follow-up question later via email?
		\item Snowball: One last thing: I'm trying to learn as much as possible about the use of the Sentence Risk Assessment Instrument. I was wondering if you might be able to put me in touch with other judges to talk about the tool.
	\end{itemize}
	
\end{enumerate}


\section{Code Table}
\label{sec:code_table}

\small
\centering

\begin{supertabular}{  l } 
	
	\textbf{Sentencing Practice} \\ 
	
	\hline

	Comments on using the sentencing guidelines\\
	Describe their sentencing process as different or unusual/better\\
	Emphasize importance of community safety\\
	Consider recidivism to be important in sentencing decision\\
	Consider seriousness of next offense more important than raw recidivism \\
	Comments on judicial discretion\\
	Mention importance of getting many cases through/efficiency \vspace{.2em}\\

	\textbf{PSI ordering behavior} 		\\	\hline

	Ordering a PSI is correlated with the seriousness of the case\\
	Always orders PSIs for trial cases\\
	Never/almost never order PSIs\\
	Explicitly say that PSIs aren't helpful\\
	Say that PSIs are helpful in more serious cases\\
	Concerned about how slow generating a PSI is\\
	Order PSI to clarify `stale' records\vspace{.2em}\\

	\textbf{Information and training}	\\
	
	\hline

	Received training or attended CJE about the tool\\
	Never received training or attended CJE about tool\\
	Heard about tool primarily in email/documentation\\
	Generally not attending/paying much attention to CJEs\\
	CJEs are helpful\vspace{.2em}\\
	
	\textbf{Familiarity and misconceptions}	\\		\hline
	
	Misconceptions about what the tool was or how it worked\\
	Wasn't sure where the risk assessment information was presented\\
	Embarrassment or shame about lack of awareness of tool\\
	
	\textbf{Use of the tool}\\			\hline

	Do not use/pay attention to the risk assessment tool\\
	Pay attention to risk assessment tool\\
	Tool has never changed decision to order PSI\\
	Tool is not used in their county\vspace{.2em}\\

	\textbf{Desires and concerns}\\
	\hline

	PSI should be generated earlier\\
	Desire access to more information not provided by this tool\\
	Mention racial bias concerns with risk assessment\\
	Concern that tool ignores defendant's humanity\\
	Explicitly say that risk assessment tool isn't helpful\\
	Think judges infer risk level from the tool\\
	Complain about unintuitiveness of SGS/guidelines form \\
	Risk assessment isn't doing anything new\vspace{.2em}\\
	
	\textbf{Broader attitudes}\\
	\hline

	Positive view of risk assessment more broadly\\
	Skeptical or negative view of risk assessment\\
	Risk assessment useful in other areas of CJ but not for judges\\
	Risk assessment might be useful for less experienced judges\\
	Purpose of risk assessment is to increase consistency\\
	Purpose of risk assessment is to increase efficiency\vspace{.2em}\\

	\hline
\end{supertabular}


\end{document}